# Heat Transfer in the Flow of a Cold, Axisymmetric Jet over a Hot Sphere

**Jian-Jun SHU**[*] and **Graham WILKS**

School of Mechanical & Aerospace Engineering, Nanyang Technological University,
50 Nanyang Avenue, Singapore 639798

**ABSTRACT**: The heat transfer characteristics of thin film flow over a hot sphere resulting from a cold vertical jet of liquid falling onto the surface has been investigated. The underlying physical features have been illustrated by numerical solutions of high accuracy based on the modified Keller box method. The solutions for film thickness distribution are good agreement with those obtained approximately by using the Pohlhausen integral momentum technique and observed experimentally by using water as working fluid, thus providing a basic confirmation of the validity of the results presented.

## 1 Introduction

Comprehensive literature reviews on theoretical and experimental studies of the flow and heat transfer characteristics for jet impingement on surfaces were made by Thome [1] and Ribatski & Jacobi [2] to reveal fully the past developments. Although many different jet characteristics were discussed, a great majority of such studies were dealt with jet impingement on a flat surface. Jet impingement on curved surfaces is very common in engineering applications. Nevertheless, too few research works considered the question.

After having examined the jet impingement on a flat plate [3], it is natural to move on to a closer inspection of the jet impingement on a sphere. The flow of a thin liquid film under gravity over a sphere occurs frequently in a variety of industrial heat-transfer applications, such as, heat exchange in coating operations [4,5], high precision wetted spheres [6], multiple sphere absorbers [7] and hydrofluidization systems [8]. In order to understand the operations and, in particular, the efficiency of these processes, it is important to have a detailed study of such flows. The assessment of heat transfer characteristics in such settings is based on Nusselt theory. Unfortunately the absence of inertia in the theory leads to the prediction of zero heat transfer at the upper generator of an inundated sphere. The work that follows in part addresses this inconsistency. Any underlying methodology of solution may indicate how best to incorporate inertia into a detailed assessment of the heat transfer characteristics of jet impingement on a sphere. Analyses of such flows [9-12] fail to distinguish between the effects of a thin, high speed jet as compared to a thick, low speed jet when each give rise to a common flow rate. Mitrovic [13] indicates that such a distinguishing capability is called for the development of theoretical models. However previous studies of thin film flow over a sphere were confined purely to the hydrodynamic problem. Gyure & Krantz [14] used a perturbation analysis for low Reynolds numbers. Gribben [15] obtained an approximation by using the Pohlhausen integral momentum technique [16], which assumed an approximate velocity profile across the thickness of the film. Hunt [17] obtained a numerical solution by using the modified Keller box method, which accommodated the outer, free boundary. Heat-transfer characteristics of the flow have not been considered.

---

[*] Correspondence should be addressed to Jian-Jun SHU,
  Email address: mjjshu@ntu.edu.sg; Phone number: (Singapore) +65 6790 4459.

In this paper, the heat transfer in the flow of a cold, axisymmetric jet over a hot sphere is investigated. The accurate and comprehensive numerical solutions for both velocity and temperature distributions are obtained by using the modified Keller box method. The main idea of the Keller box method [18] is to replace higher derivatives by first derivatives through the introduction of additional variables, and to discretize the resultant differential system by centered-difference derivatives with a second order truncation error at midpoints of net rectangles.

**Nomenclature**

| | | | |
|---|---|---|---|
| $a$ | sphere radius $(m)$ | $x_s$ | domain-related parameter |
| $F$ | gravity-related parameter | | |
| $F_r$ | Froude number | | |
| $(f, u, v, \phi, w)$ | dimensionless variables | Greek symbols | |
| $G$ | geometry-related parameter | $\alpha$ | thickness-related parameter |
| $g$ | acceleration $(m/s^2)$ | $\beta$ | similarity-related parameter |
| $H$ | film thickness $(m)$ | $\gamma$ | curvature-related parameter |
| $h$ | dimensionless film thickness | $\kappa$ | thermal diffusivity $(m^2/s)$ |
| $N_u$ | Nusselt number | $\nu$ | kinematic viscosity $(m^2/s)$ |
| $P$ | pressure $(kg/ms^2)$ | $(\xi, \eta)$ | dimensionless coordinates |
| $P_r$ | Prandtl number | $\rho$ | density $(kg/m^3)$ |
| $p$ | dimensionless pressure | $\psi$ | stream function |
| $Q$ | flow rate $(m^3/s)$ | | |
| $R_e$ | Reynolds number | Superscript | |
| $(\theta, r), (x, Y), (x, y)$ | spherical coordinates measured by angular displacement and radial distance respectively | $-$ | dimensional analysis |
| $T$ | temperature $(K)$ | | |
| $T_w$ | wall temperature $(K)$ | Subscript | |
| $(U, V)$ | velocity components $(m/s)$ | 0 | jet |

## 2 Modelling

The problem to be examined concerns the film cooling which occurs when a cold vertically draining column strikes a hot sphere. Although a column of fluid draining under gravity is accelerated and thin at impact [19,20], it is reasonable to model the associated volume flow as a jet of uniform velocity $U_0$, uniform temperature $T_0$ and radius $H_0$ as is illustrated in Figure 1. The notation $Q = \pi H_0^2 U_0$ is introduced for the flow rate and a film Reynolds number may be defined as $R_e = \dfrac{U_0 a}{\nu}$ based on the sphere radius $a$.

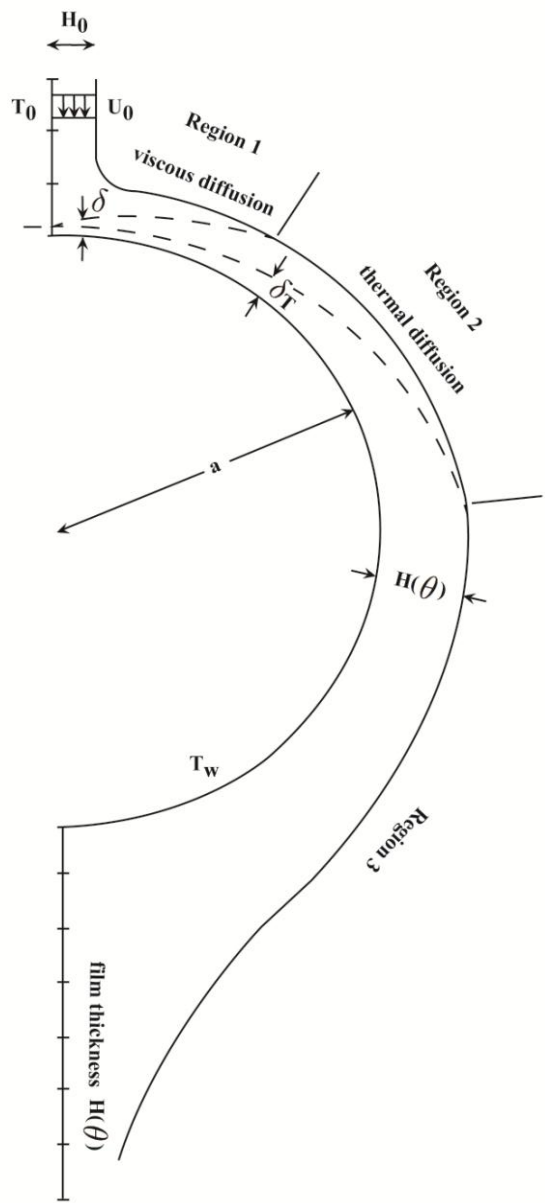

**Figure 1**: The vertical jet and resultant film for the sphere

# 3 Governing Equations

The flow under investigation has been modelled as a steady, axisymmetric flow of incompressible fluid. For the large Reynolds number case in which the flow is confined to a thin film, quantities vary much less in the streamwise direction than in the wall normal direction. In the absence of viscous dissipation, the resultant governing equations expressing conservation of mass, momentum and energy become very similar to the boundary layer equations and are consequently

$$\frac{1}{\sin\theta}\frac{\partial}{\partial\theta}(V_\theta \sin\theta) + \frac{1}{r}\frac{\partial}{\partial r}(r^2 V_r) = 0 \qquad (1)$$

$$\frac{V_\theta}{r}\frac{\partial V_\theta}{\partial\theta} + v_r \frac{\partial V_\theta}{\partial r} + \frac{V_r V_\theta}{r} = g\sin\theta - \frac{1}{\rho r}\frac{dP}{d\theta}$$
$$+ \nu\left[\frac{1}{r^2}\frac{\partial}{\partial r}\left(r^2 \frac{\partial V_\theta}{\partial r}\right) + \frac{1}{r^2 \sin\theta}\frac{\partial}{\partial\theta}\left(\sin\theta \frac{\partial V_\theta}{\partial\theta}\right) + \frac{2}{r^2}\frac{\partial V_r}{\partial\theta} - \frac{V_\theta}{r^2 \sin^2\theta}\right] \qquad (2)$$

$$\frac{v_\theta}{r}\frac{\partial T}{\partial\theta} + v_r \frac{\partial T}{\partial r} = \kappa\left[\frac{1}{r^2}\frac{\partial}{\partial r}\left(r^2 \frac{\partial T}{\partial r}\right) + \frac{1}{r^2 \sin\theta}\frac{\partial}{\partial\theta}\left(\sin\theta \frac{\partial T}{\partial\theta}\right)\right] \qquad (3)$$

where $(V_\theta, V_r)$ are velocity components associated with spherical coordinates $(\theta, r)$ measured by the angular displacement from the top of the sphere and the radial distance from the centre of the sphere respectively.

In the specified physical setting, the equations are to be solved subject to the following conditions.

(i) The no slip boundary condition at the wall requires that

$$V_\theta = V_r = 0 \quad \text{on } r = a, \ 0 \le \theta \le \pi. \qquad (4)$$

(ii) The temperature at the wall is assumed constant as $T_w$, say

$$\text{i.e.} \quad T = T_w \quad \text{on } r = a, \ 0 \le \theta \le \pi. \qquad (5)$$

(iii) On the free surface of the film, prescribed by $r = a + H(\theta)$, the shearing stress may be assumed negligible and consequently

$$r\frac{\partial}{\partial r}\left(\frac{V_\theta}{r}\right) + \frac{1}{r}\frac{\partial V_r}{\partial\theta} = 0 \quad \text{at } r = a + H(\theta), \ 0 \le \theta \le \pi. \qquad (6)$$

(iv) Similarly, in a film cooling environment such as water surrounded by air, it may be assumed that there is negligible heat flux on the free surface and hence that

$$\frac{\partial T}{\partial r}=0 \quad \text{at} \quad r=a+H(\theta),\ 0\leq\theta\leq\pi. \tag{7}$$

(v) Once an overall flow rate $Q=\pi H_0^2 U_0$ has been prescribed, a conservation of volume flow constraint at any given $\theta$ station leads to the condition

$$2\pi\sin\theta\int_a^{a+H(\theta)} rV_\theta(\theta,r)dr = \text{constant} = \pi H_0^2 U_0 \quad \text{for} \quad 0\leq\theta\leq\pi. \tag{8}$$

Under the assumption that the film thickness remains thin relative to a characteristic horizontal dimension, a boundary layer treatment of the equations leads to significant simplification.

The following non-dimensional variables are introduced

$$x=\theta,\ \overline{Y}=\frac{R_e^{\frac{1}{2}}(r-a)}{a},\ \overline{H}(x)=\frac{R_e^{\frac{1}{2}}H(\theta)}{a},$$

$$\overline{U}=\frac{V_\theta}{U_0},\ \overline{V}=\frac{R_e^{\frac{1}{2}}V_r}{U_0},\ \phi=\frac{T-T_w}{T_0-T_w},\ p=\frac{P}{\rho U_0^2}. \tag{9}$$

The big O notation, $O$, is used to describe the limiting behavior of a function for a very small argument. In the limit $R_e\to+\infty$ with $x$ remaining $O(1)$ and after neglecting terms of $O\left(\frac{1}{R_e^{\frac{1}{2}}}+\frac{H_0^2}{a^2}\right)$ compared with unity, the following equations are obtained

$$\frac{\partial}{\partial x}(\overline{U}\sin x)+\frac{\partial}{\partial\overline{Y}}(\overline{V}\sin x)=0 \tag{10}$$

$$\overline{U}\frac{\partial\overline{U}}{\partial x}+\overline{V}\frac{\partial\overline{U}}{\partial\overline{Y}}=\frac{1}{F_r}\sin x-\frac{dp}{dx}+\frac{\partial^2\overline{U}}{\partial\overline{Y}^2} \tag{11}$$

$$P_r\left(\overline{U}\frac{\partial\phi}{\partial x}+\overline{V}\frac{\partial\phi}{\partial\overline{Y}}\right)=\frac{\partial^2\phi}{\partial\overline{Y}^2} \tag{12}$$

where $P_r=\frac{\nu}{\kappa}$ is the Prandtl number and $F_r=\frac{U_0^2}{ag}$ is the Froude number based on the jet velocity on its surface. In common with standard boundary layer theory, $\frac{dp}{dx}$ implies that the pressure across the film remains constant. In the absence of external pressure gradients and with zero shear assumed on the free surface, the pressure term in (11) is identically zero.

The boundary conditions now read
(i) $\quad\quad\quad \overline{U}=\overline{V}=0 \quad \text{on} \quad \overline{Y}=0,\ 0\leq x\leq\pi$ (13)
(ii) $\quad\quad\quad \phi=0 \quad \text{on} \quad \overline{Y}=0,\ 0\leq x\leq\pi$ (14)

(iii) $\qquad \dfrac{\partial \overline{U}}{\partial \overline{Y}} = 0 \quad \text{at} \quad \overline{Y} = \overline{H}(x), \ 0 \le x \le \pi$ (15)

(iv) $\qquad \dfrac{\partial \phi}{\partial \overline{Y}} = 0 \quad \text{at} \quad \overline{Y} = \overline{H}(x), \ 0 \le x \le \pi$ (16)

(v) $\qquad \displaystyle\int_{0}^{\overline{H}(x)} \overline{U} \, d\overline{Y} = \dfrac{R_e^{\frac{1}{2}} H_0^2}{2a^2 \sin x} \quad \text{for} \ 0 \le x \le \pi.$ (17)

## 4 Numerical Solutions

The continuity equation (10) can be eliminated by introducing a stream function $\psi$ defined by
$$\overline{U} = \dfrac{1}{\sin x} \dfrac{\partial \psi}{\partial \overline{Y}}, \ \overline{V} = -\dfrac{1}{\sin x} \dfrac{\partial \psi}{\partial x}. \tag{18}$$
Owing to the geometry, $\overline{H}(x)$ is singular at $x = 0$ and $x = \pi$. To remove this singularity, $y$ and $h(x)$ are introduced and given by
$$y = \sin x \, \overline{Y}, \ h(x) = \sin x \, \overline{H}(x). \tag{19}$$
Substituting equations (18) and (19) into (10)-(17) gives
$$\dfrac{\partial^3 \psi}{\partial y^3} + \dfrac{1}{F_r \sin x} = \left(\dfrac{1}{\sin^2 x}\right)\left(\dfrac{\partial \psi}{\partial y}\dfrac{\partial^2 \psi}{\partial x \partial y} - \dfrac{\partial \psi}{\partial x}\dfrac{\partial^2 \psi}{\partial y^2}\right) \tag{20}$$

$$\dfrac{\partial^2 \phi}{\partial y^2} = \left(\dfrac{P_r}{\sin^2 x}\right)\left(\dfrac{\partial \psi}{\partial y}\dfrac{\partial \phi}{\partial x} - \dfrac{\partial \psi}{\partial x}\dfrac{\partial \phi}{\partial y}\right) \tag{21}$$

subject to boundary conditions
$$\psi = 0, \ \dfrac{\partial \psi}{\partial y} = 0, \ \phi = 0 \quad \text{on} \ y = 0, \ 0 \le x \le \pi \tag{22}$$

$$\psi = \dfrac{R_e^{\frac{1}{2}} H_0^2}{2a^2}, \ \dfrac{\partial^2 \psi}{\partial y^2} = 0, \ \dfrac{\partial \phi}{\partial y} = 0 \quad \text{on} \ y = h(x), \ 0 \le x \le \pi \tag{23}$$

$$h = \dfrac{R_e^{\frac{1}{2}} H_0^2}{2a^2}, \ \psi = y, \ \phi = 1 \quad \text{on} \ x = 0, \ 0 < y \le \dfrac{R_e^{\frac{1}{2}} H_0^2}{2a^2} \tag{24}$$

where the initial condition (24) appears due to the original initial condition
$$H = -a + \sqrt{a^2 + \dfrac{H_0^2}{\sin \theta}}, \ V_\theta = U_0, \ T = T_0 \quad \text{on} \ \theta = 0, \ a < r \le \sqrt{a^2 + \dfrac{H_0^2}{\sin \theta}}. \tag{25}$$

The main idea of the Keller box method [18] is to replace higher derivatives by first derivatives through the introduction of additional variables, and to discretize the resultant differential system by centered-difference derivatives with a second order truncation error at midpoints of net rectangles. Some progress [21-25] has been made in investigating various problems of a liquid jet impinging on a solid surface. In anticipation of the use of a Keller box method and its attractive extrapolation features the differential system (20)-(24) is re-cast as the following first order system

$$\frac{\partial \psi}{\partial y} = u$$

$$\frac{\partial u}{\partial y} = \bar{v}$$

$$\frac{\partial \bar{v}}{\partial y} = -\frac{1}{F_r \sin x} + \left(\frac{1}{\sin^2 x}\right)\left(u\frac{\partial u}{\partial x} - \bar{v}\frac{\partial \psi}{\partial x}\right) \quad (26)$$

$$\frac{\partial \phi}{\partial y} = \bar{w}$$

$$\frac{\partial \bar{w}}{\partial y} = \left(\frac{P_r}{\sin^2 x}\right)\left(u\frac{\partial \phi}{\partial x} - \bar{w}\frac{\partial \psi}{\partial x}\right)$$

whose boundary conditions are

$$\psi = 0, \; u = 0, \; \phi = 0 \quad \text{on} \; y = 0, \; 0 \leq x \leq \pi$$

$$\psi = \frac{R_e^{\frac{1}{2}} H_0^2}{2a^2}, \; \bar{v} = 0, \; \bar{w} = 0 \quad \text{on} \; y = h(x), \; 0 \leq x \leq \pi \quad (27)$$

$$h = \frac{R_e^{\frac{1}{2}} H_0^2}{2a^2}, \; \psi = y, \; \phi = 1 \quad \text{on} \; x = 0, \; 0 < y \leq \frac{R_e^{\frac{1}{2}} H_0^2}{2a^2}.$$

If these equations were used as the basis of solution there would be strong comparisons between the associated algorithm and that developed in [26]. However the expectation of an initial Blasius boundary layer within the film can be assimilated into the solution scheme by further transformations. The underlying methodology of solution nevertheless remains the same. In each case the discretisation is aimed at generating a simultaneous system of nonlinear equations which can be solved by Newton iteration.

According to the non-dimensional transformation, the boundary layer thickness grows like $x^{\frac{3}{2}}$ for small $x$ in the $y$ direction.

The following coordinate transformation, what simultaneously maps the film thickness onto the unit interval and removes the Blasius singularity at the origin, is introduced

$$x = \xi^{\frac{2}{3}}, \; y = \frac{\xi \eta h}{\xi + 1 - \eta}. \quad (28)$$

The dependent variables are transformed as

$$\psi = \frac{\xi}{\xi + 1 - \eta} f, \; \bar{v} = \frac{\xi + 1 - \eta}{\xi} v, \; \bar{w} = \frac{\xi + 1 - \eta}{\xi} w. \quad (29)$$

The equations to be solved now read

$$f_\eta = \frac{(1 + \xi) h u}{\xi + 1 - \eta} - \frac{f}{\xi + 1 - \eta}$$

$$u_\eta = \frac{(1+\xi)hv}{\xi+1-\eta}$$

$$v_\eta = \frac{v}{\xi+1-\eta} - \frac{\xi^2(1+\xi)h}{F_r(\xi+1-\eta)^3 \sin\left(\xi^{\frac{2}{3}}\right)} - \frac{3(1-\eta)\xi^{\frac{4}{3}}(1+\xi)hfv}{2(\xi+1-\eta)^4 \sin^2\left(\xi^{\frac{2}{3}}\right)} + \frac{3\xi^{\frac{7}{3}}(1+\xi)h}{2(\xi+1-\eta)^3 \sin^2\left(\xi^{\frac{2}{3}}\right)}\left(uu_\xi - vf_\xi\right)$$

(30)

$$\phi_\eta = \frac{(1+\xi)hw}{\xi+1-\eta}$$

$$w_\eta = \frac{w}{\xi+1-\eta} - \frac{3P_r(1-\eta)\xi^{\frac{4}{3}}(1+\xi)hfw}{2(\xi+1-\eta)^4 \sin^2\left(\xi^{\frac{2}{3}}\right)} + \frac{3P_r\xi^{\frac{7}{3}}(1+\xi)h}{2(\xi+1-\eta)^3 \sin^2\left(\xi^{\frac{2}{3}}\right)}\left(u\phi_\xi - wf_\xi\right)$$

subject to

$$f = 0,\ u = 0,\ \phi = 0 \quad \text{on } \eta = 0,\ 0 \le \xi \le \pi^{\frac{3}{2}}$$

$$f = \frac{R_e^{\frac{1}{2}} H_0^2}{2a^2},\ v = 0,\ w = 0 \quad \text{on } \eta = 1,\ 0 \le \xi \le \pi^{\frac{3}{2}} \quad (31)$$

$$h = \frac{R_e^{\frac{1}{2}} H_0^2}{2a^2},\ f = f_0(\eta),\ \phi = \phi_0(\eta) \quad \text{on } \xi = 0,\ 0 < \eta \le 1$$

where the initial profiles $f_0(\eta)$ and $\phi_0(\eta)$ are found by putting $\xi = 0$ and $h = \frac{R_e^{\frac{1}{2}} H_0^2}{2a^2}$ into (30) and solving subject to conditions $f = u = \phi = 0$ at $\eta = 0$ and $u = 1$, $\phi = 1$ at $\eta = 1$.

The parabolic system of equations and boundary conditions (30)-(31) has been solved by marching in the $\xi$-direction based on a modification of the Keller box method. A non-uniform grid is placed on the domain $\xi \ge 0$, $0 \le \eta \le 1$ and the resultant difference equations are solved by Newton iteration. Solutions are obtained on different sized grids and Richardson's extrapolation used to produce results of high accuracy.

It is worth to mention to this end that the general equations describing parabolic free boundary problems arising from thin film flows [26] are

$$\frac{\partial^3 \psi}{\partial y^3} + F(x) = G^2(x)\left(\frac{\partial \psi}{\partial y}\frac{\partial^2 \psi}{\partial x \partial y} - \frac{\partial \psi}{\partial x}\frac{\partial^2 \psi}{\partial y^2}\right) \quad (32)$$

$$\frac{\partial^2 \bar{\phi}}{\partial y^2} = P_r G^2(x)\left(\frac{\partial \psi}{\partial y}\frac{\partial \bar{\phi}}{\partial x} - \frac{\partial \psi}{\partial x}\frac{\partial \bar{\phi}}{\partial y}\right) \quad (33)$$

subject to boundary conditions

$$\psi = 0,\ \frac{\partial \psi}{\partial y} = 0,\ \bar{\phi} = 0 \quad \text{on } y = 0,\ 0 \le x \le x_s \quad (34)$$

$$\psi = \gamma, \quad \frac{\partial^2 \psi}{\partial y^2} = 0, \quad \frac{\partial \bar{\phi}}{\partial y} = 0 \quad \text{at} \quad y = \bar{h}(x), \ 0 \le x \le x_s \tag{35}$$

$$\bar{h} = \gamma, \ \psi = y, \ \bar{\phi} = 1 \quad \text{on} \quad x = 0, \ 0 < y \le \gamma. \tag{36}$$

Accordingly the variables are transformed from ($x, y$) to ($\xi, \eta$) by using

$$x = \xi^\alpha, \quad y = \frac{\xi \eta \bar{h}}{\xi + 1 - \eta}. \tag{37}$$

The dependent variables are transformed as

$$\psi = \frac{\xi}{\xi + 1 - \eta} f, \ \bar{u} = \frac{1}{(1+\xi)^\beta} u, \ \bar{v} = \frac{\xi + 1 - \eta}{\xi(1+\xi)^{2\beta}} v, \ \bar{\phi} = \phi, \ \bar{w} = \frac{\xi + 1 - \eta}{\xi(1+\xi)^\beta} w, \ \bar{h} = (1+\xi)^\beta h. \tag{38}$$

Here, for the sphere in this paper, the relevant physical parameters in the parabolic system of general equations, boundary conditions and coordinate transformation (32)-(38) describing free boundary problems arising from thin film flows should be chosen as the gravity-related parameter $F(x) = \frac{1}{F_r \sin x}$, the geometry-related parameter $G(x) = \frac{1}{\sin x}$, the domain-related parameter $x_s = \pi$, the curvature-related parameter $\gamma = \frac{R_e^{\frac{1}{2}} H_0^2}{2a^2}$, the thickness-related parameter $\alpha = \frac{2}{3}$ and the similarity-related parameter $\beta = 0$. A full account of the numerical method and the details of implementation are beyond the scope of this paper and is reported separately [26]. The detailed numerical method procedure for this case is fully discussed in [27]. The solution scheme was successfully tested against previously reported results [28-33].

## 5 Results

A typical run has a coarse grid of dimensions $60 \times 48$ in the ($\xi, \eta$) domain with each cell being divided into *1*, *2*, *3* and *4* sub-cells respectively. Because of the coordinate singularity at $\xi = 0$, $\eta = 1$, a non-uniform grid is employed and given by $\xi = \bar{\xi}^{1.75}$, $\eta = 1 - (1 - \bar{\eta})^{1.5}$ where $\bar{\xi}$ and $\bar{\eta}$ are uniform. When $\Delta \bar{\xi} \equiv \frac{1}{59} \pi^{\frac{6}{7}}$ and $\Delta \bar{\eta} \equiv \frac{1}{47}$, this gives $\Delta \xi \sim 0.004$ and $\Delta \eta \sim 0.003$ near the singularity, which is sufficiently small to give good accuracy. From the convergence of the extrapolation process, the absolute error is $6 \times 10^{-7}$. A typical set of numerical data is presented in Table.

**Table**: Film thickness, free surface velocity and temperature for the sphere with $F_r = 1$, $\gamma = 0.5$ and $P_r = 2$

| $x$ | film thickness $\bar{h}(x)$ | free surface velocity $\bar{u}(x, \bar{h}(x))$ | free surface temperature $\bar{\phi}(x, \bar{h}(x))$ |
|---|---|---|---|
| 0.000 | 0.500 | 1.000 | 1.000 |
| 0.218 | 0.585 | 1.023 | 1.000 |
| 0.396 | 0.678 | 1.063 | 1.000 |
| 0.587 | 0.775 | 0.997 | 0.964 |
| 0.786 | 0.891 | 0.864 | 0.831 |
| 0.994 | 1.007 | 0.756 | 0.654 |
| 1.208 | 1.086 | 0.695 | 0.484 |
| 1.427 | 1.128 | 0.667 | 0.343 |
| 1.595 | 1.141 | 0.658 | 0.260 |
| 1.823 | 1.138 | 0.659 | 0.180 |
| 1.996 | 1.122 | 0.667 | 0.139 |
| 2.290 | 1.072 | 0.695 | $9.760 \times 10^{-2}$ |
| 2.590 | 0.994 | 0.744 | $7.779 \times 10^{-2}$ |
| 3.018 | 0.861 | 0.823 | $7.022 \times 10^{-2}$ |
| $\pi$ | 0.836 | 0.831 | $7.013 \times 10^{-2}$ |

In Figure 2, the numerical solution for the film thickness distribution over the sphere is compared with Gribben's approximation [15] for $F_r = 1$ and $\gamma = 0.5$. The agreement is seen to be surprisingly good. Figures 3-9 depict the flavour of the numerical results. Figures 3-4, Figures 5-6 and Figures 7-9 show film thickness, free surface velocity and free surface temperature respectively for various cases.

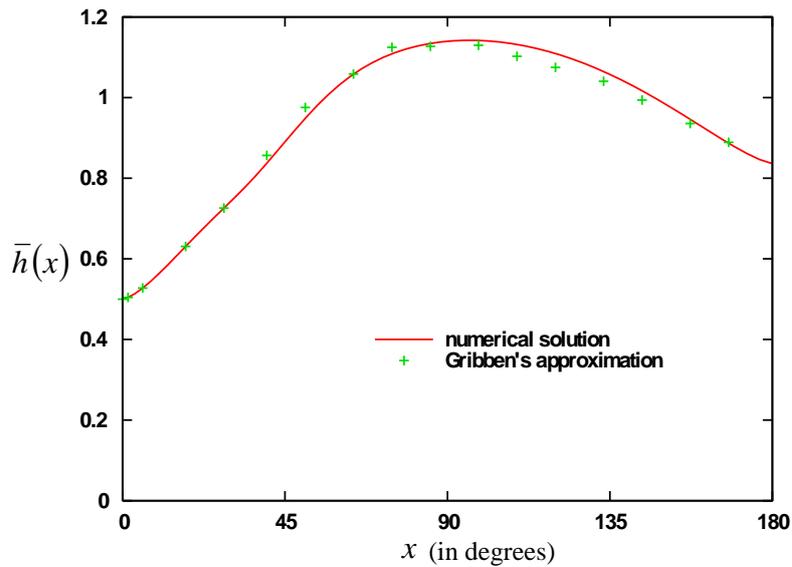

**Figure 2**: Film thickness for the numerical solution and Gribben's approximation at $F_r = 1$ and $\gamma = 0.5$

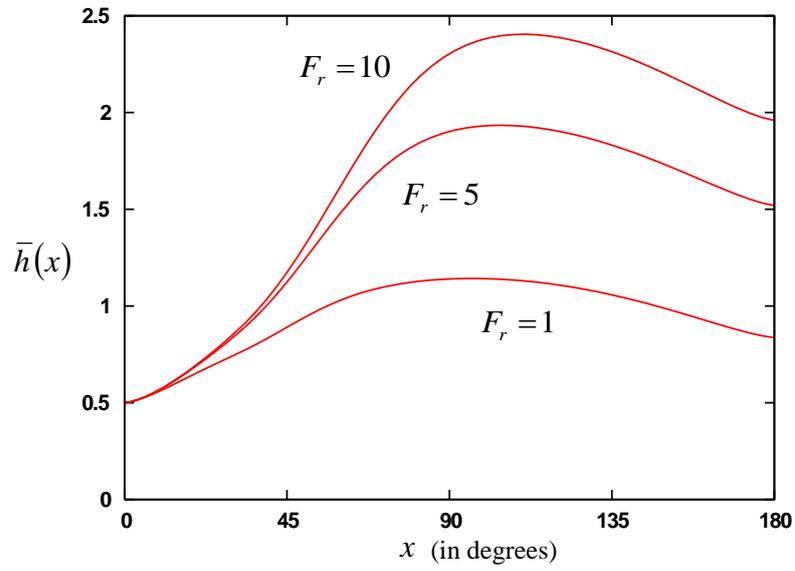

**Figure 3**: Film thickness for various Froude numbers at $\gamma = 0.5$

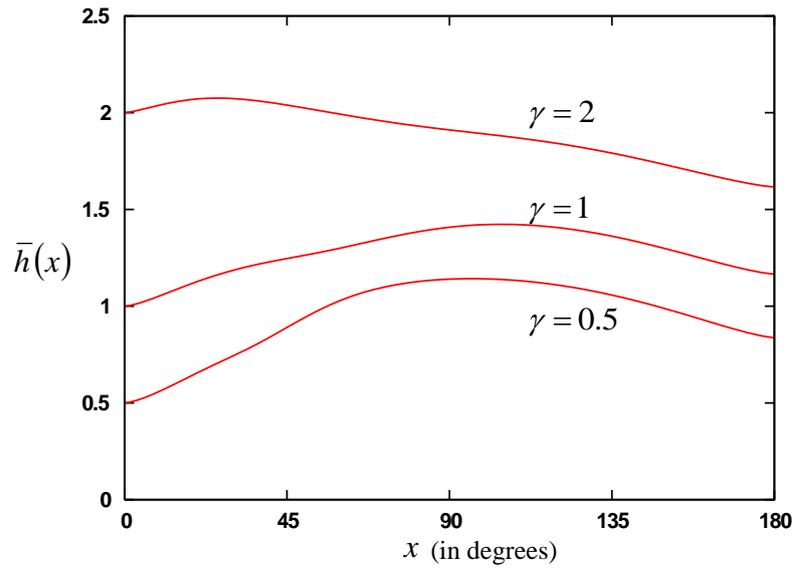

**Figure 4**: Film thickness for various values of the parameter $\gamma$ at $F_r = 1$

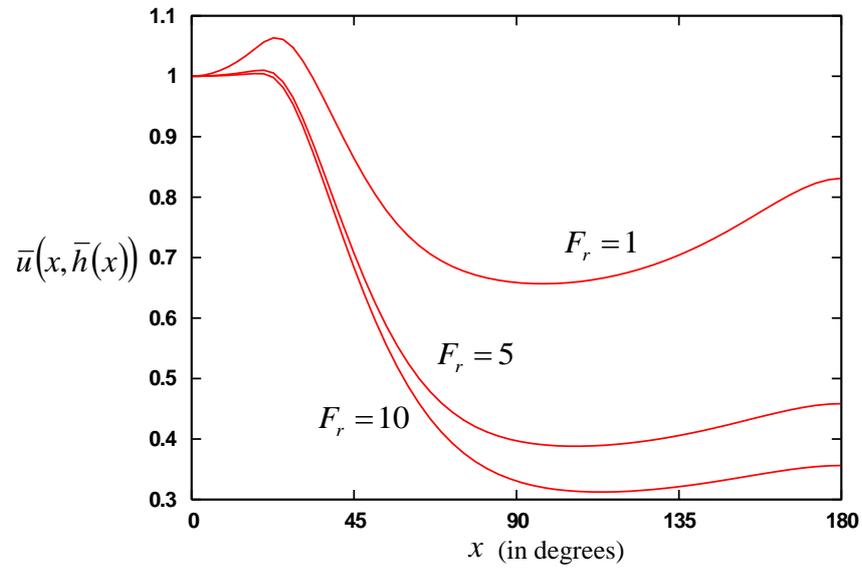

**Figure 5**: Free surface velocity for various Froude numbers at $\gamma = 0.5$

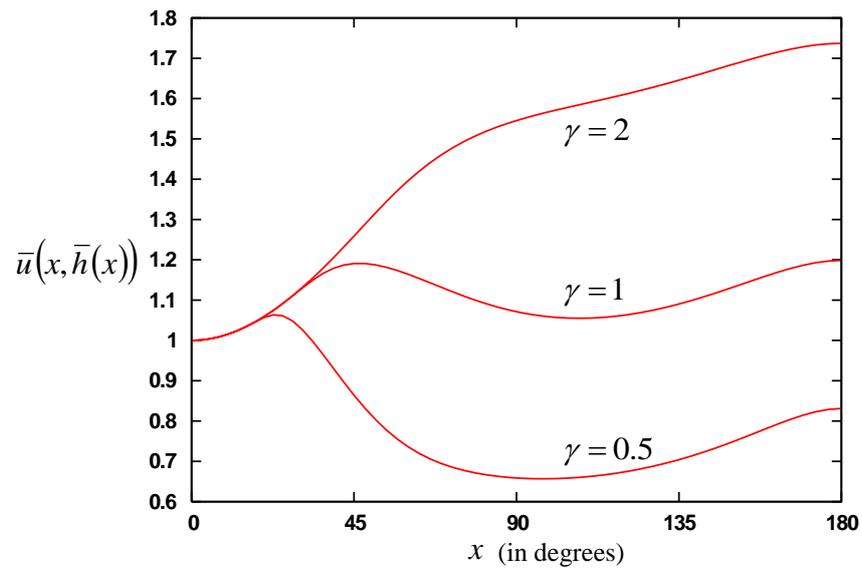

**Figure 6**: Free surface velocity for various values of the parameter $\gamma$ at $F_r = 1$

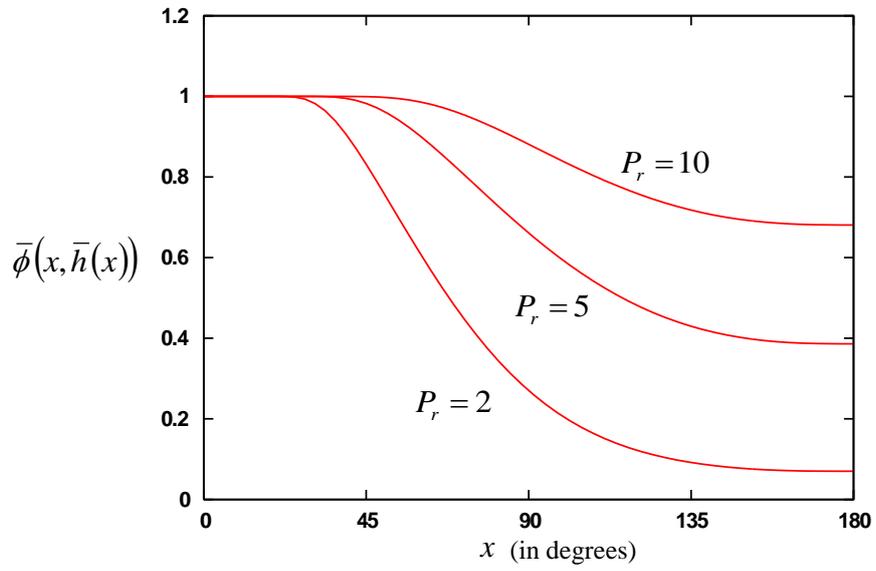

**Figure 7**: Free surface temperature for various Prandtl numbers at $F_r = 1$ and $\gamma = 0.5$

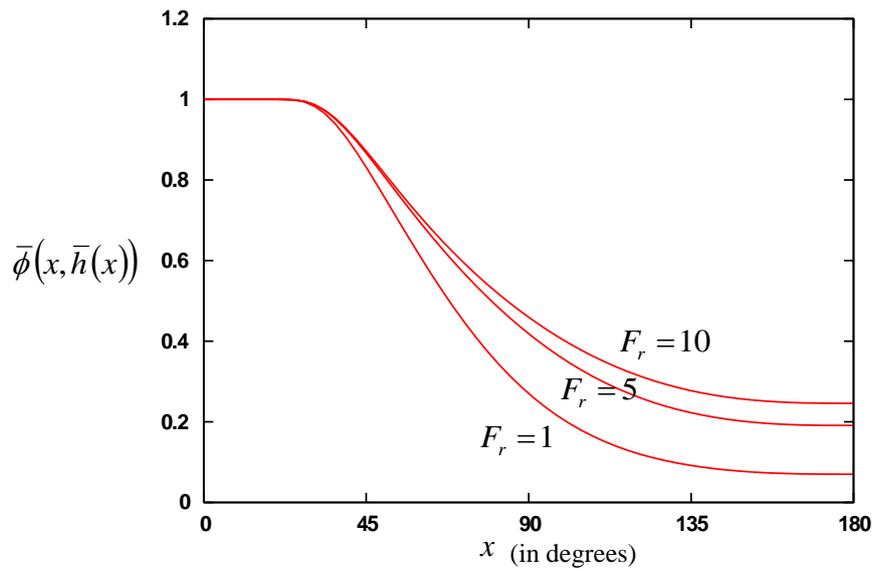

**Figure 8**: Free surface temperature for various Froude numbers at $\gamma = 0.5$ and $P_r = 2$

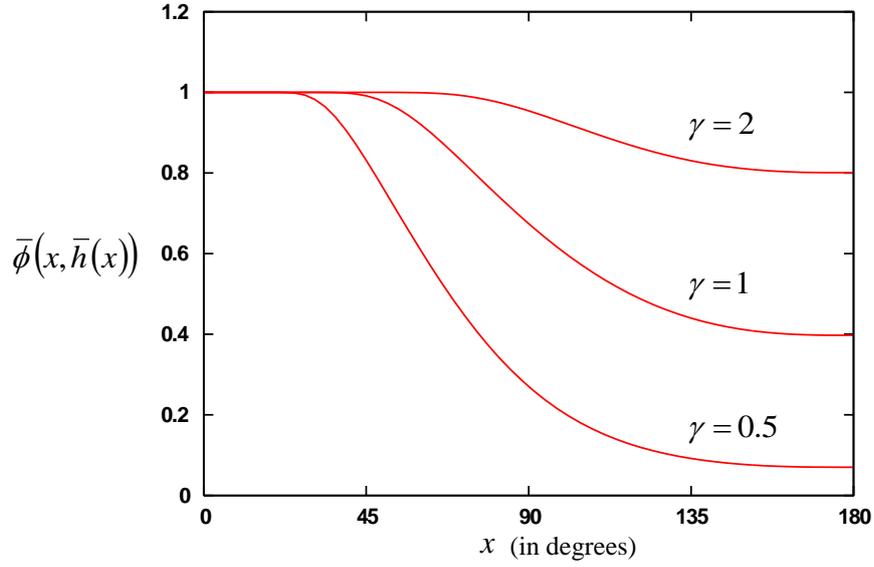

**Figure 9**: Free surface temperature for various values of the parameter $\gamma$ at $F_r = 1$ and $P_r = 2$

It is worth to mention that the author's previous published results on a flat plate [3] can be regarded as a limiting case of the present new ones on a sphere, when the focus is only on the near field around jet impingement point $x \to 0$ with negligible gravity effect $F_r \to \infty$ and negligible curvature effect $\gamma \to 0.5$. This can be evidenced by making the comparison between Figures 3, 5 and 8 for the upper hemispherical surface $0 \leq x \leq \frac{\pi}{2}$ at $F_r \geq 5$ and the counterpart figures in [3] (collectively as shown in Figure 10 for the sake of easy reference), but for the lower hemispherical surface $\frac{\pi}{2} \leq x \leq \pi$ at $F_r < 5$, the characteristics are totally different between two cases.

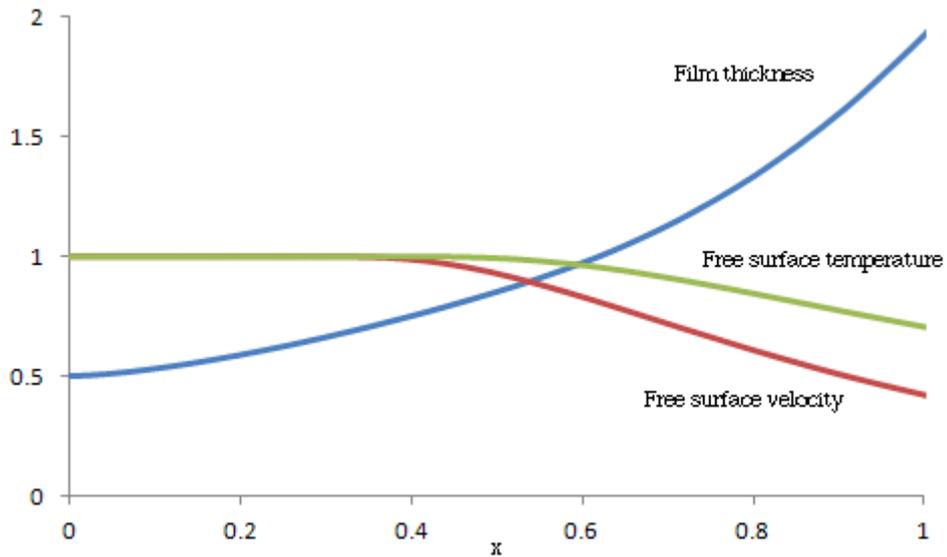

**Figure 10**: Film thickness, free surface temperature and free surface velocity for the flat plate case [3]

For the sphere case, the velocity of the flow is controlled by two opposing forces, viscosity trying to slow it down (less heat transfer) and gravity trying to speed it up (more heat transfer). The viscous component of force affecting the flow is greatest near $x = \frac{\pi}{2}$ and least near $x = 0$ and $x = \pi$. Figure 5 show that at $F_r = 1$, free surface velocity initially has a slight increase, followed by a sharp decrease as viscosity starts to dominate, and finally a gradual increase is observed as the bottom of the sphere is approached. This corresponds to the up-down situation for the film thickness in Figures 2-4, and the decline at the different rates for the free surface temperature in Figures 7-9. This phenomenon was observed experimentally [34] by using water as working fluid for jet impingement Reynolds number ranging from the order of $10^4$ to $10^6$. As $F_r$ decreases, the effect of gravity increases and hence the thin film thickness, high velocity and low temperature appear corresponding to the small $F_r$ values. As $\gamma$ decreases, the amount of fluid in the impinging jet decreases and the ensuing film becomes thinner. The effect of viscosity increases and hence the low velocity and low temperature appear corresponding to the small $\gamma$ values. Where $\gamma$ is greater than a certain value, e.g., $\gamma = 2$ for $F_r = 1$, Figure 6 shows no decrease for the velocity over the sphere. As $P_r$ increases, the temperature decrease becomes more gradual.

It is worth to mention to this end that the most significant film cooling design factor is the heat transfer across the film. This can be presented most conveniently by introduction of the Nusselt number defined as

$$N_u = \left.\frac{\partial \phi}{\partial y}\right|_{y=0} = \frac{x^{\frac{3}{2}} + 1}{x^{\frac{3}{2}}} w\Big|_{\eta=0}. \tag{39}$$

A typical result of the Nusselt number, $N_u$, for the sphere with $F_r = 1$, $\gamma = 0.5$ and $P_r = 2$ is illustrated in Figure 11. It is shown that the Nusselt number, $N_u$, increases monotonically as the rate of $x^{\frac{3}{2}}$ as approaching to the top of the sphere. This is in contrast to models based solely on a balance of viscous and gravitational terms, which necessarily predict zero Nusselt number at the top of the sphere. In contrast to the classical Nusselt theory, the film inertia may generate significant heat transfer (high Nusselt number) at the top of the sphere.

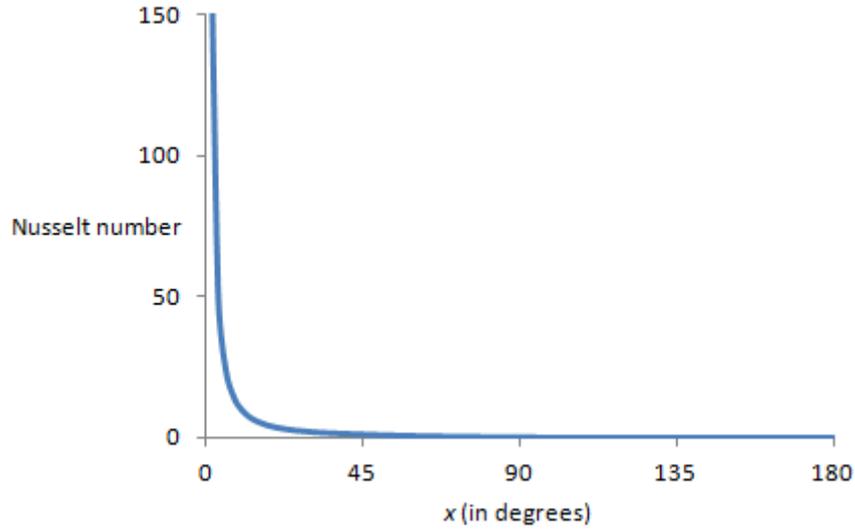

**Figure 11**: Nusselt number for the sphere with $F_r = 1$, $\gamma = 0.5$ and $P_r = 2$

## 6 Concluding Remarks

A detailed examination of a jet over a sphere has been performed. The accurate and comprehensive numerical solutions for establishing the flow and heat transfer characteristics of a cold axi-symmetric jet over a hot sphere have been presented by modifying the Keller box method to accommodate the outer, free boundary. The gross features of such flows have been illustrated over a range of representative parameter values. These indicate the underlying features of the developing film thickness, velocity and temperature distributions. The comparison with the experimental observation [34] by using water as working fluid for jet impingement Reynolds number ranging from the order of $10^4$ to $10^6$ is to indicate that the current numerical solution may be carried over with confidence to the sphere inundation problem, thus providing a basis of comparison with Mitrovic's experimental results [13]. The work also provides the basis for re-assessing condensation drainage and inundation flows; recognizing that in contrast to Nusselt theory, the inertia of inundating film may generate significant heat transfer at the top of a flooded sphere. In particular such an implementation of the present method is likely to yield significant heat transfer at the top of an inundated sphere. This is in contrast to models based solely on a balance of viscous and gravitational terms, which necessarily predict zero heat transfer at the upper generator. In a practical setting, appropriate parameter values may be evaluated and the design characteristics readily identified from the numerical solutions. In practice, it is not obvious that uniform wetting of the sphere would occur. Instabilities may distort or even disrupt such a uniform distribution, that is, $U_0$ may be not considered as average velocity rather than uniform velocity. A full account of non-linear flow instability analysis based on perturbation method is very tedious and beyond the scope of this paper. Nevertheless for a given overall flow rate, the model may represent a valuable first approximation to the aggregate properties of the flow.

## References

[1] Thome, J.R., 1999, Falling film evaporation: State-of-the-art review of recent work, *Journal of Enhanced Heat Transfer*, **6**(2-4), 263-277.
[2] Ribatski, G. and Jacobi, A.M., 2005, Falling-film evaporation on horizontal tubes - A critical review, *International Journal of Refrigeration-Revue Internationale du Froid*, **28**(5), 635-653.
[3] Shu, J.-J. and Wilks, G., 2008, Heat transfer in the flow of a cold, axisymmetric vertical liquid jet against a hot, horizontal plate, *Journal of Heat Transfer-Transactions of the ASME*, **130**(1), 012202.
[4] Sjösten, J., Golriz, M.R., Nordin, A. and Grace, J.R., 2004, Effect of particle coating on fluidized-bed heat transfer, *Industrial & Engineering Chemistry Research*, **43**(18), 5763-5769.
[5] Sjösten, J., Golriz, M.R. and Grace, J.R., 2006, Further study on the influence of particle coating on fluidized bed heat transfer, *International Journal of Heat and Mass Transfer*, **49**(21-22), 3800-3806.
[6] Tzeng, S.-H. and Yang, S.-A., 2007, Second law analysis and optimization for film-wise condensation from downward flowing vapors onto a sphere, *Heat and Mass Transfer*, **43**(4), 365-369.
[7] Weislogel, M.M. and Chung J.N., 1991, Experimental investigation of condensation heat transfer in small arrays of PCM-filled spheres, *International Journal of Heat and Mass Transfer*, **34**(1), 31-45.
[8] Peralta, J.M., Rubiolo, A.C. and Zorrilla, S.E., 2009, Design and construction of a hydrofluidization system. Study of the heat transfer on a stationary sphere, *Journal of Food Engineering*, **90**(3), 358-364.
[9] Solan, A. and Zfati, A., 1974, Heat transfer in laminar flow of a liquid film on a horizontal cylinder, *Heat Transfer 1974, Proceedings of the Fifth International Heat Transfer Conference*, Keidanrenkaikan Building, Tokyo, Japan, **II**, 90-93.
[10] Parken, W.H. and Fletcher, L.S., 1978, Heat transfer in thin liquid films flowing over horizontal tubes, *Proceedings of the Sixth International Heat Transfer Conference*, Toronto, Canada, 415-420.
[11] Rogers, J.T., 1981, Laminar falling film flow and heat-transfer characteristics on horizontal tubes, *Canadian Journal of Chemical Engineering*, **59**(2), 213-222.
[12] Andberg, J.W. and Vliet, G.C., 1986, Absorption of vapors into liquid films flowing over cooled horizontal tubes, *Heat Transfer 1986, Proceedings of the Eighth International Heat Transfer Conference*, San Francisco, CA.
[13] Mitrovic, J., 1986, Influence of tube spacing and flow-rate on heat-transfer from a horizontal tube to a falling liquid-film, *Heat Transfer 1986, Proceedings of the Eighth International Heat Transfer Conference*, San Francisco, CA.
[14] Gyure, D.C. and Krantz, W.B., 1983, Laminar film flow over a sphere, *Industrial & Engineering Chemistry Fundamentals*, **22**(4), 405-410.
[15] Gribben, R.J., 1987, Laminar film flow over a sphere at high Reynolds number, *Mathematical Engineering Industry*, **1**(4), 279-288.
[16] Rosenhead, L., 1988, *Laminar boundary layers*, Dover Publications.
[17] Hunt, R., 1989, The numerical solution of parabolic free boundary problems arising from thin film flows, *Journal of Computational Physics*, **84**(2), 377-402.
[18] Keller, H.B., 1970, A new difference scheme for parabolic problems, In: *Numerical solutions of partial differential equations*, **2**, (J. Bramble *et al*., eds.), 327-350, Academic Press, New York.
[19] Shu, J.-J., 2004, Impact of an oblique breaking wave on a wall, *Physics of Fluids*, **16**(3), 610-614.
[20] Shu, J.-J., 2004, Slamming of a breaking wave on a wall, *Physical Review E*, **70**(6), 066306.

[21] Shu, J.-J. and Wilks, G., 1996, Heat transfer in the flow of a cold, two-dimensional vertical liquid jet against a hot, horizontal plate, *International Journal of Heat and Mass Transfer*, **39**(16), 3367-3379.

[22] Zhao, Y.H., Masuoka, T., Tsuruta, T. and Ma, C.-F., 2002, Conjugated heat transfer on a horizontal surface impinged by circular free-surface liquid jet, *JSME International Journal Series B-Fluids and Thermal Engineering*, **45**(2), 307-314.

[23] Shu, J.-J., 2004, Microscale heat transfer in a free jet against a plane surface, *Superlattices and Microstructures*, **35**(3-6), 645-656.

[24] Guerra, D.R.S., Su, J. and Freire, A.P.S., 2005, The near wall behavior of an impinging jet, *International Journal of Heat and Mass Transfer*, **48**(14), 2829-2840.

[25] Rice, J., Faghri, A. and Cetegen, B., 2005, Analysis of a free surface film from a controlled liquid impinging jet over a rotating disk including conjugate effects, with and without evaporation, *International Journal of Heat and Mass Transfer*, **48**(25-26), 5192-5204.

[26] Shu, J.-J. and Wilks, G., 1995, An accurate numerical method for systems of differentio-integral equations associated with multiphase flow, *Computers & Fluids*, **24**(6), 625-652.

[27] Shu, J.-J. and Wilks, G., 2009, Heat transfer in the flow of a cold, two-dimensional draining sheet over a hot, horizontal cylinder, *European Journal of Mechanics B-Fluids*, **28**(1), 185-190.

[28] Shu, J.-J. and Wilks, G., 1995, Mixed-convection laminar film condensation on a semi-infinite vertical plate, *Journal of Fluid Mechanics*, **300**, 207-229.

[29] Shu, J.-J. and Pop, I., 1997, Inclined wall plumes in porous media, *Fluid Dynamics Research*, **21**(4), 303-317.

[30] Shu, J.-J. and Pop, I., 1998, Transient conjugate free convection from a vertical flat plate in a porous medium subjected to a sudden change in surface heat flux, *International Journal of Engineering Science*, **36**(2), 207-214.

[31] Shu, J.-J. and Pop, I., 1998, On thermal boundary layers on a flat plate subjected to a variable heat flux, *International Journal of Heat and Fluid Flow*, **19**(1), 79-84.

[32] Shu, J.-J. and Pop, I., 1999, Thermal interaction between free convection and forced convection along a vertical conducting wall, *Heat and Mass Transfer*, **35**(1), 33-38.

[33] Shu, J.-J., 2012, Laminar film condensation heat transfer on a vertical, non-isothermal, semi-infinite plate, *Arabian Journal for Science and Engineering*, **37**(6), 1711-1721.

[34] Hu, G.X. and Zhang, L.X., 2007, Experimental and numerical study on heat transfer with impinging circular jet on a convex hemispherical surface, *Heat Transfer Engineering*, **28**(12), 1008-1016.